\documentclass[aps, prc, twocolumn, tightenlines, letterpaper, preprintnumbers, floatfix, longbibliography, nofootinbib,superscriptaddress,10pt]{revtex4-2}

\usepackage[utf8]{inputenc}
\usepackage{amsmath}
\usepackage{graphicx}
\usepackage{dsfont}
\usepackage{physics}
\usepackage{placeins}
\usepackage{amssymb}
\usepackage{mathtools}
\usepackage{bbm}
\usepackage{amssymb}
\usepackage{tabularx}
\usepackage[colorlinks=true, allcolors=blue]{hyperref}
\usepackage{multirow}
\usepackage[T1]{fontenc}
\usepackage{bm}
\usepackage{xspace}
\usepackage[dvipsnames]{xcolor}
\usepackage[normalem]{ulem}
\usepackage{diagbox}
\usepackage{soul}
\usepackage{nccmath}
\usepackage{lipsum}
\usepackage{cancel}
\usepackage{bm}
\usepackage{algorithm2e}
\usepackage{tikz}
\usetikzlibrary{quantikz2}

\newcolumntype{Y}{>{\centering\arraybackslash}X}

\usepackage{orcidlink}

\definecolor{lavenderindigo}{rgb}{0.58, 0.34, 0.92}

\usepackage{stackengine}
\stackMath
\newcommand\tenq[2][1]{
 \def\useanchorwidth{T}%
  \ifnum#1>1%
    \stackunder[0pt]{\tenq[\numexpr#1-1\relax]{#2}}{\scriptscriptstyle\sim}%
  \else%
    \stackunder[1pt]{#2}{\scriptscriptstyle\sim}%
  \fi%
}

\makeatletter  
\renewcommand\onecolumngrid{
\do@columngrid{one}{\@ne}%
\def\set@footnotewidth{\onecolumngrid}
\def\footnoterule{\kern-6pt\hrule width 1.5in\kern6pt}%
}
\makeatother    

\begin{document}

\begin{figure}
\vskip -1.cm
\leftline{\includegraphics[width=0.15\textwidth]{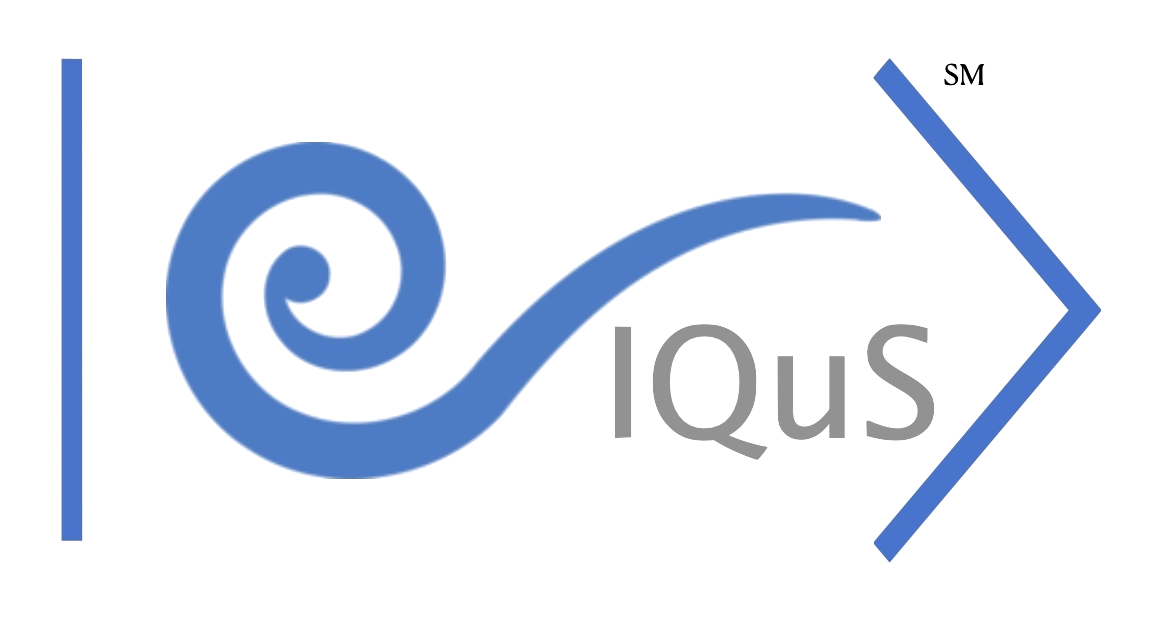}}
\end{figure}

\title{Exclusive Scattering Channels from Entanglement Structure in Real-Time Simulations}

\author{Nikita A. Zemlevskiy\,\orcidlink{0000-0002-0794-2389}}
\email{zemlni@uw.edu}
\affiliation{InQubator for Quantum Simulation (IQuS), Department of Physics, University of Washington, Seattle, WA 98195, USA}

\preprint{IQuS@UW-21-121}
\date{\today}

\begin{abstract}
\noindent 
A scattering event in a quantum field theory is a coherent superposition of all processes consistent with its symmetries and kinematics.
While real-time simulations have progressed toward resolving individual channels, existing approaches rely on knowledge of the asymptotic particle wavefunctions.
This work introduces an experimentally inspired method to isolate scattering channels in Matrix Product State simulations based on the entanglement structure of the late-time wavefunction.
Schmidt decompositions at spatial bipartitions of the post-scattering state identify elastic and inelastic contributions, enabling deterministic detection of outgoing particles of specific species.
This method may be used in settings beyond scattering and is applied to detect heavy particles produced in a collision in the one-dimensional Ising field theory.
Natural extensions to quantum simulations of other systems and higher-order processes are discussed.
\end{abstract}
\maketitle


\section{Introduction}
\label{sec:intro}
\noindent
High-energy collisions in which energy is converted into matter through particle production generically result in final states that are coherent superpositions of many possible configurations.
The late-time wavefunction after a scattering event encodes the possible products of the collision and their respective amplitudes.
Decomposing the full superposition into individual channels, i.e., resolving the exclusive final states rather than summing inclusively over all configurations, is a prerequisite for extracting branching ratios and reconstructing channel-specific kinematics.
Collider facilities such as RHIC and the LHC are built on the principle that channel-specific observables carry far more information than inclusive measurements~\cite{ParticleDataGroup:2024cfk}.
For instance, the relative abundances of color-singlet configurations after a heavy-ion collision can reveal the mechanism of hadronization~\cite{Rafelski:1982pu,ALICE:2016fzo}.
The same principle applies across energy scales, from nuclear decay channels relevant to nucleosynthesis pathways~\cite{Burbidge:1957vc} to quantum chemical reactions where several processes occur in superposition~\cite{Kassal:2010xwg,Manthe2016Smatrix}.
Extracting comparable information from real-time simulations requires methods to resolve exclusive final states.

While classical analysis of strongly-coupled scattering is largely limited to static observables, first-principles real-time simulations are possible on a sufficiently powerful quantum computer.
Following the demonstrations of state preparation algorithms~\cite{Farrell:2024fit,Davoudi:2024wyv} and first simulations of elastic collisions~\cite{Zemlevskiy:2024vxt,Hite:2025pvb,Chai:2023qpq,Chai:2025kbi}, inelastic scattering has recently been probed in simulations on quantum computers~\cite{Schuhmacher:2025ehh,Farrell:2025nkx}.
Classical simulations of collisions in one-dimensional systems with moderate growth of entanglement are possible using Matrix Product States (MPS), which faithfully represent the post-scattering state with a tractable bond dimension.
While such simulations provide access to the inclusive post-scattering wavefunction, extracting the contributions of individual channels from this many-body state is nontrivial. 

This work shows that the entanglement structure, revealed through the Schmidt decomposition, provides a natural way to resolve these channels.
Individual channels are identified through the tensor product factorization at bipartitions separating components from different channels.
This method leverages spatial separations in the state to dissect the full superposition into orthogonal components, each corresponding to a distinct scattering outcome, with Schmidt coefficients that directly yield the channel probabilities.
Physical meaning has been ascribed to Schmidt vectors in various contexts, for instance in string breaking and hadronization~\cite{Grieninger:2026bdq,Florio:2024aix}, in characterizing topological features~\cite{Li:2008kda,Fidkowski:2010nhf}, and identifying quasiparticle excitations~\cite{Zauner-Stauber:2018gqr,Cocchiarella:2025mtv}.
In the context of collisions, entanglement and quantum complexity measures provide a powerful lens on scattering and particle production dynamics~\cite{MISHIMA2004371,Robin:2025ymq,Martin:2025hzm,Cheng:2025zaw,Mendez:2024wqn,Quinta:2023ink,Zagirdinova:2022awg,Hentschinski:2022rsa,Cervera-Lierta:2017tdt,Hida:2009jdg,Harshman:2007omw,Giorgi:2006lse,Yuasa:2006faz,Lamata:2006au,Lombardi:2006tzp,Harshman:2006sec,Peschanski:2019yah,Bai:2022hfv,Kowalska:2024kbs,Hales:2022osm,PhysRevA.73.052313,Kouzakov:2019hbt,Karlsson_2021,Sou:2025tyf,Tkachev:2024iuj,Gargalionis:2025iqs,Beane:2018oxh,Robin:2024oqc}.
Recent work has shown that the entanglement at a spatial bipartition characterizes the nature of the process~\cite{Jha:2024jan,Milsted:2020jmf,Pavesic:2025nwm,Papaefstathiou:2024zsu,Belyansky:2023rgh,Barata:2025rjb,Barata:2025hgx,Rigobello:2021fxw,Van_Damme_2021,Bennewitz:2024ixi}. 
In particular, it is known that the late-time state after an inelastic collision is more entangled than the elastic case.
This work reveals a correspondence between the presence of additional reaction channels and the entanglement structure of the post-scattering state.

This correspondence is used to analyze the final state of scattering simulations by isolating channels based on the spatial structure of the late-time wavefunction.
This method is demonstrated in simulations of inelastic collisions in the Ising field theory, where channel isolation enables the extraction of particle masses from individual scattering outcomes and provides confirmation of heavy particle production.
While the present work focuses on scattering, the approach extends naturally to other settings where distinct physical processes occur in superposition.

\section{Isolating scattering channels}
\label{sec:method}
\begin{figure}
    \centering
    \includegraphics[width=\linewidth]{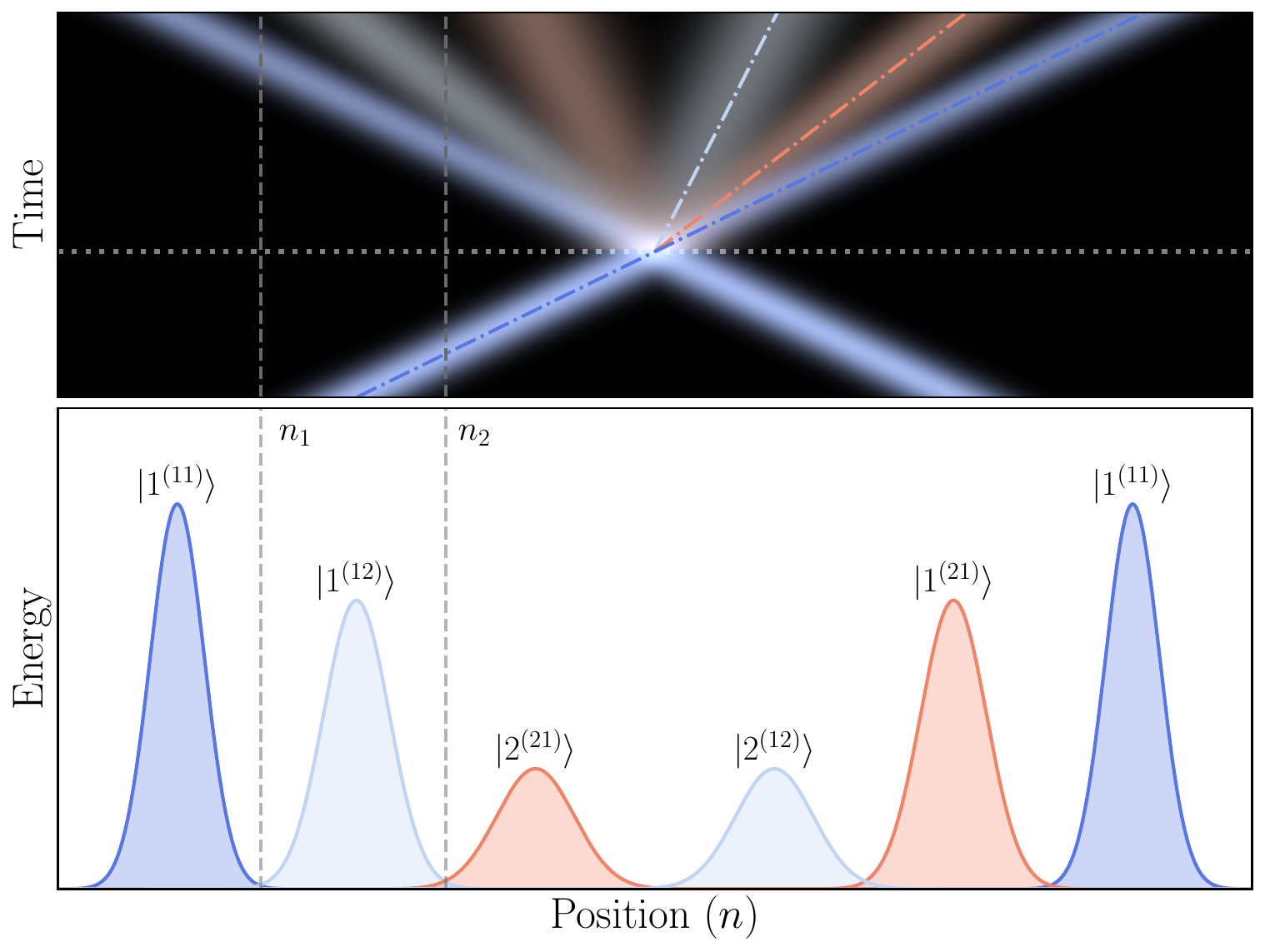}
    \caption{{\it Channel isolation in the post-inelastic scattering state.}
    Top: the energy density as a function of position and time with the $|11\rangle$ (dark blue), $|12\rangle$ (light blue), and $|21\rangle$ (orange) states distinguished.
    The dotted line marks the collision time and the dot-dashed lines show the trajectories of the outgoing particles in each state.
    Bottom: the energy density as a function of position for the final state in the heatmap in the top panel. 
    The superscripts on the individual particles label the states they belong to.
    The dashed lines show the positions $n_1$ and $n_2$ where bipartitions can be made to isolate $|11\rangle$, $|12\rangle$, and $|21\rangle$.
    }
    \label{fig:late_time_schematic}
\end{figure}

\noindent
In regimes where both elastic and inelastic two-particle scattering is possible, these processes happen in superposition. 
The method to isolate these channels is general and only requires spatial separation of collinear outgoing particles.
For concreteness, consider a theory in one spatial dimension with two stable particles, $|1\rangle$ and $|2\rangle$, such that $m_1<m_2$.
In a high-energy collision, kinetic energy can be converted into mass, forming a heavier outgoing particle.
Such particle production is possible in a collision of two $|1\rangle$ with momentum $k_i>k_\text{thr}$, where $k_\text{thr}$ is the threshold momentum at which the total energy is equal to $m_1+m_2$.\footnote{Working in the center-of-momentum frame, this threshold corresponds to $\sqrt{s} \geq m_1 + m_2$ in terms of the Lorentz-invariant Mandelstam variable $s = (E_1 + E_2)^2 - (k_1 + k_2)^2$. For simplicity, higher-order processes such as 2-3 scattering are not considered.} 
The scattering process of two $|1\rangle$ particles with $k_i>k_\text{thr}$ is
\begin{equation}
\begin{aligned}
   |1(k_i)\rangle |1(-k_i)\rangle \ \to \ \alpha &|1^{(11)}(-k_i)\rangle |1^{(11)}(k_i)\rangle \\ + \ \beta &|1^{(12)}(-k_f)\rangle |2^{(12)}(k_f)\rangle \\ + \ \gamma &|2^{(21)}(-k_f)\rangle |1^{(21)}(k_f)\rangle \\ 
   \equiv \ \ &|f\rangle \ ,
   \label{eq:production}
\end{aligned}
\end{equation}
where the first entry represents an outgoing particle traveling left, and the second entry corresponds to a particle moving right.
The superscripts label the state each particle belongs to and $\beta=\gamma$ by parity symmetry.
Throughout this work, 11 and 12 refer to the elastic and inelastic channels, and $|11\rangle$, $|12\rangle$, and $|21\rangle$ represent the states that contribute to the late-time wavefunction.
For theories where energy depends nontrivially on momentum, excitations generally have different group velocities $v_{1,2}(k)=dE_{1,2}(k)/dk$.
Wavepackets corresponding to different particle species therefore separate in space over time, allowing different scattering channels to be distinguished by the presence or absence of excitations in particular spatial regions.
An example of the late-time energy density corresponding to Eq.~\eqref{eq:production} is shown in Fig.~\ref{fig:late_time_schematic}.

Previous work has determined particle content by constructing projectors onto localized states with given quantum numbers~\cite{Jha:2024jan,Van_Damme_2021}. 
Entanglement considerations are central to MPS simulations and allow leveraging spatial separation between channels with different outgoing velocities.
The key ingredient of the method in this work is the Schmidt decomposition across a bipartition between collinear components of different channels, which forces the wavefunction into a sum of tensor product states. 
By finding the basis that diagonalizes correlations between two regions, the Schmidt decomposition naturally yields the spatially-separated channels. 
When all wavepackets are spatially separated, the states in Eq.~\eqref{eq:production} occupy distinct regions of space.
Since only $|1^{(11)}\rangle$ has support to the left of $n_1$ in Fig.~\ref{fig:late_time_schematic}, it is orthogonal to $|1^{(12)}\rangle$ and $|2^{(21)}\rangle$.
The Schmidt decomposition at $n_1$ therefore isolates the fast-moving elastic components from the slower inelastic ones,
\begin{equation}
    |f\rangle \ = \ \alpha|1^{(11)}\rangle\otimes|1^{(11)}\rangle + \beta|\text{vac}\rangle \otimes|\psi_{12}\rangle \ ,
    \label{eq:proj_n1}
\end{equation}
where $|\text{vac}\rangle$ schematically denotes the vacuum to the left of the cut and $|\psi_{12}\rangle =|1^{(12)}\rangle |2^{(12)}\rangle + |2^{(21)}\rangle |1^{(21)}\rangle$.
The tensor product structure at $n_l$ is made explicit through $\otimes$.
Retaining the second term in this decomposition projects onto the wavefunction of the inelastic channel.\footnote{This projection is implemented in MPS through the singular value decomposition (SVD) at the bipartition. 
The state of interest is obtained by setting the singular values corresponding to all other Schmidt vectors to zero.}
Since only the left-moving $|1^{(12)}\rangle$ has support between $n_1$ and $n_2$, a cut at $n_2$ on the second term in Eq.~\eqref{eq:proj_n1} separates the two branches of the inelastic channel, 
\begin{equation}
    |\psi_{12}\rangle \ = \ |1^{(12)}\rangle \otimes |2^{(12)}\rangle + |\text{vac}\rangle \otimes |2^{(21)}\rangle|1^{(21)}\rangle \ .
\end{equation}
The exclusive states corresponding to this decomposition are shown by different colors in the bottom panel of Fig.~\ref{fig:late_time_schematic}.

From the separation into the elastic and inelastic channels it is straightforward to extract the branching ratio for particle production.
The Schmidt coefficients give the probabilities for each process:
\begin{equation}
    P(11) \ = \ |\alpha|^2 \ , \quad
    P(12) \ = \ 2|\beta|^2 \ ,
    \label{eq:branching_ratios}
\end{equation}
from which exclusive cross sections can be extracted.
This result is obvious from Eq.~\eqref{eq:production} but easily obtained from this decomposition.

The presence of a heavy particle in $|f\rangle$ is confirmed by computing the masses of excitations in the exclusive final states through the dispersion relation together with the momenta of the excitations.
Momentum may be measured with several different techniques and is discussed in App.~\ref{app:momentum}.
In this work, the velocity of local observables is matched to the group velocities of different excitations to determine their momenta, and solve for their masses from the dispersion relations.

This method is exact in the limit where the outgoing particles belonging to different channels traveling in the same direction are well-separated. 
This requires initial states of sufficiently large, collimated wavepackets, as well as sufficiently late times.
In practice, the combined effects of wavepacket spreading and finite simulation time lead to small residual mixing between channels and corresponding uncertainties in the extracted probabilities.
These approximations can be quantified and systematically reduced by increasing the momentum-space resolution of the simulation.

\section{Channel Isolation After Particle Production in Ising Field Theory }
\label{sec:results}
\noindent
Despite arising from a simple nearest-neighbor spin interaction, the Ising field theory supports a rich spectrum and nontrivial phase structure, making it an ideal setting for real-time simulations of nonperturbative phenomena.
Recent work has used it to study confinement and meson dynamics~\cite{Kormos:2016osj,Surace:2020ycc}, thermalization~\cite{Jaschke_2019}, and false-vacuum decay ~\cite{Milsted:2020jmf,Pavesic:2025nwm,borla2026microscopicdynamicsfalsevacuum}.
This work investigates particle production in an inelastic scattering process in the Ising model,
\begin{align}
\hat{H}  \ &=\  -  \sum_{n=0}^{L-1}\left [ \frac{1}{2}\left (\hat{Z}_{n-1}\hat{Z}_{n}+\hat{Z}_n\hat{Z}_{n+1}\right ) +  g_x \hat{X}_n + g_z\hat{Z}_n \right ] \nonumber \\
&\equiv \ \sum_{n=0}^{L-1}\hat{H}_n \ , \label{eq:h_ising}
\end{align}
which corresponds to a quantum field theory (QFT) when tuned near criticality (i.e., ${g_x\approx1,\ g_z\approx0,\ L\to\infty}$).
Throughout this work, the couplings $g_x=1.25,\ g_z=0.15$ and system size $L=400$ are used.
With these couplings, the theory has two stable particles, $|1\rangle$ and $|2\rangle$.
Trotter time evolution is used with MPS to simulate the scattering of two wavepackets of $|1\rangle$ particles, with $k_i=0.36\pi>k_\text{thr}$ and width in momentum space $\sigma_k=0.059\pi$.
See App.~\ref{app:ising_details} for details on this model and App.~\ref{app:sim_details} for the simulation techniques.

\begin{figure}
    \centering
    \includegraphics[width=\linewidth]{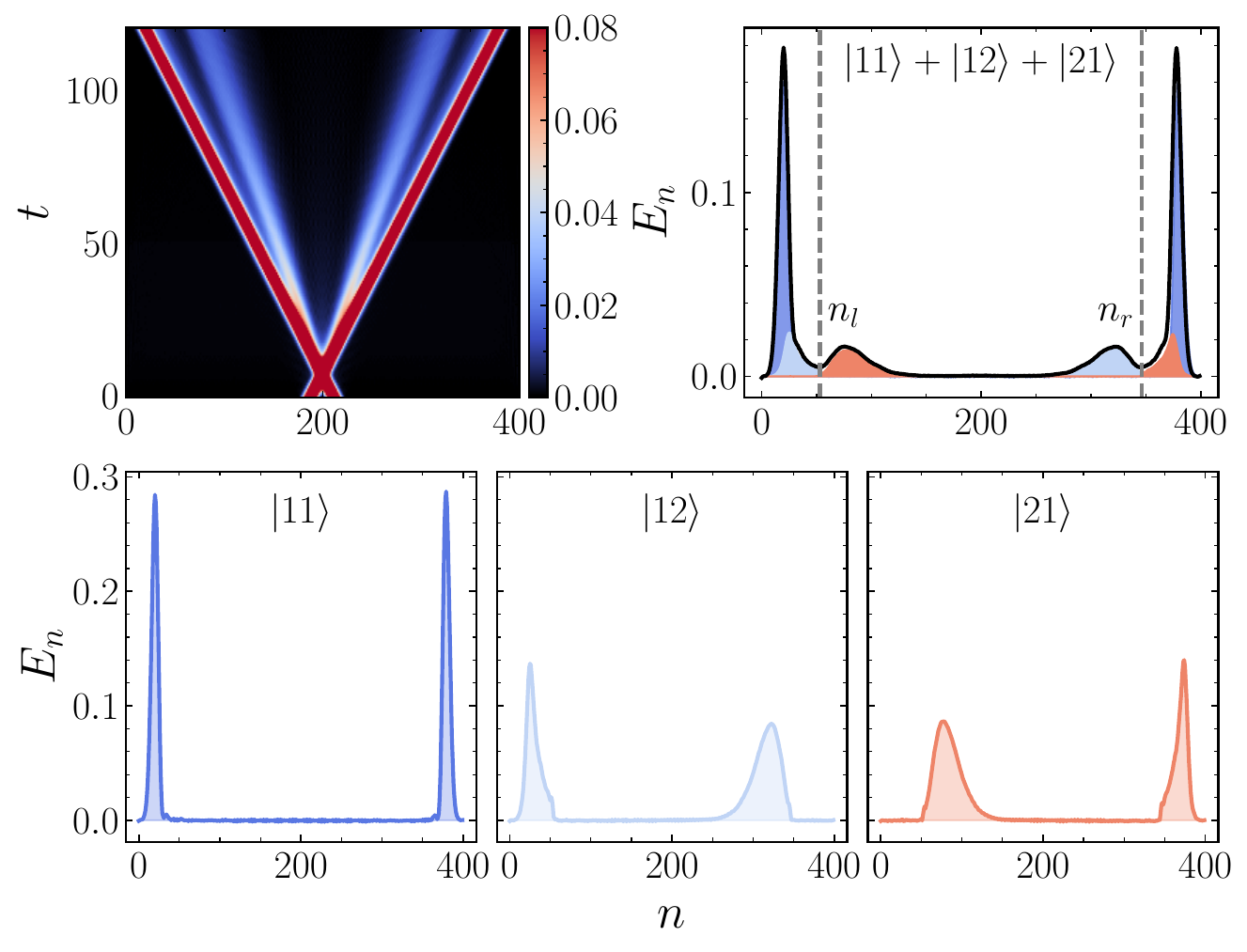}
    \caption{{\it Leading-order channel isolation after particle production.}
    Top left: the energy density $E_n$ as a function of position $n$ and time $t$ for the collision of two $|1\rangle$ particles with $k_i=\pm0.36\pi$. 
    Top right: $E_n$ corresponding to $t=120$ in the top left panel. 
    The energy density of the inclusive post-scattering state $|f\rangle$ is shown in black.
    The color shading shows the contributions of the individual states: $|11\rangle$ (dark blue), $|12\rangle$ (light blue), and $|21\rangle$ (orange).
    The positions $n_l$ and $n_r$ of the bipartitions used to isolate the states composing the elastic and inelastic channels are marked by the dashed lines.
    Bottom: $E_n$ of the exclusive states contributing to $|f\rangle$ in the top right panel.
    }
    \label{fig:channel}
\end{figure}

The top left panel of Fig.~\ref{fig:channel} shows the vacuum-subtracted energy density, 
\begin{align}
E_n \ = \ \langle f(t)|\hat{H}_n |f(t)\rangle - \langle\text{vac}|\hat{H}_n |\text{vac}\rangle\ ,
\label{eq:e_n}
\end{align}
throughout the scattering process.
As in Eq.~\eqref{eq:production}, the simulation begins with the $|1\rangle$ particles moving toward each other. 
At $t\sim8$ they collide and travel away from each other. 
After the collision, the faint tracks toward the center of the lattice represent the contributions of the slower $|2\rangle$ particle produced in the inelastic channel.

Entanglement and quantum complexity are known to be powerful indicators of dynamics in physical processes. 
For instance, entanglement entropy has been shown to track the onset of inelastic thresholds and the creation of new particles in meson collisions~\cite{Rigobello:2021fxw,Papaefstathiou:2024zsu}, the transition from fermionic Fock states to meson-like bound states during hadronization is revealed by properties of the entanglement spectrum~\cite{Florio:2023dke,Florio:2024aix}, and measures of quantum complexity expose nonlocal correlations during string breaking~\cite{Grieninger:2026bdq}.
Increased levels of entanglement have been observed in late-time states after inelastic scattering.
This work identifies the reason for this: elastic collisions contain a single dominant Schmidt vector, whereas the amount of significant Schmidt components increases with the inelasticity and the number of possible distinct processes.

\begin{figure}
    \centering
    \includegraphics[width=\linewidth]{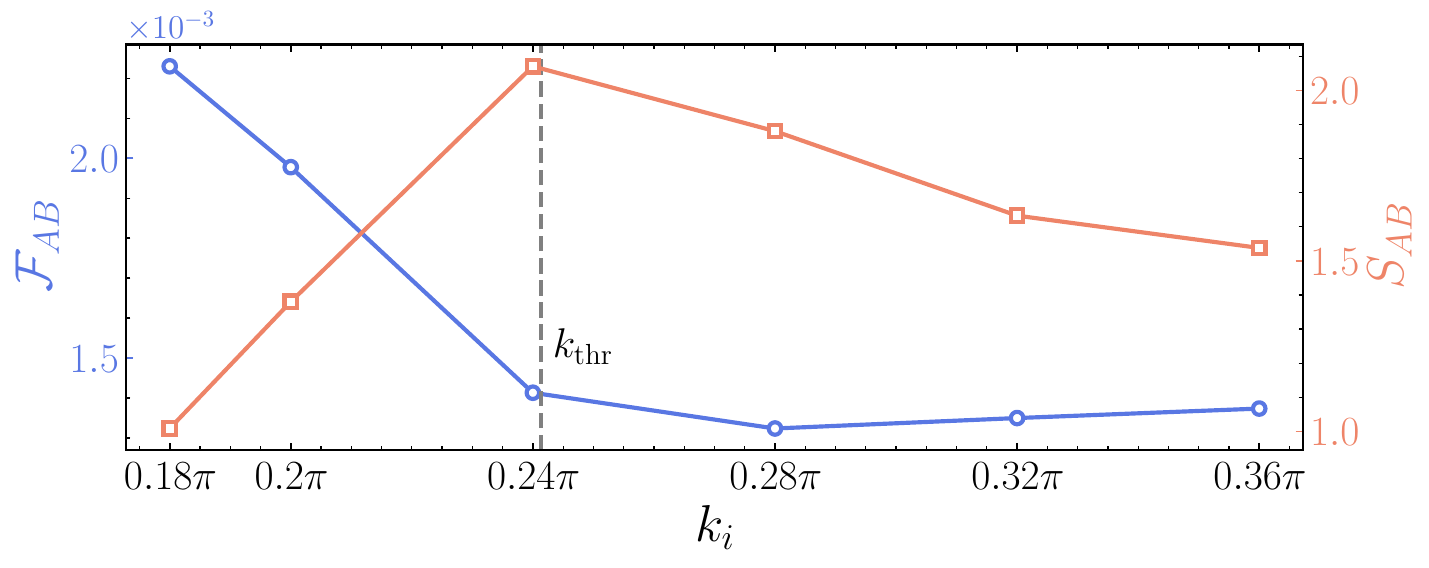}
    \caption{{\it Entanglement structure of the post-scattering state.}
    The antiflatness ${\cal F}_{AB}$ (blue) and the entanglement entropy $S_{AB}$ (orange) as a function of initial momentum $k_i$.
    The dashed line marks the threshold momentum $k_\text{thr}$ above which the process $11\to12$ is kinematically allowed.
    }
    \label{fig:af_ee}
\end{figure}

The antiflatness of the entanglement spectrum $\{\lambda_i\}$ (Schmidt coefficients squared) is measured by its variance,\footnote{Similar measures are known to bound nonstabilizerness~\cite{Tirrito:2023fnw}.} 
\begin{align}
    {\cal F}_{AB} \ = \ \sum_i \lambda_i^2 - \left(\sum_i \lambda_i\right)^2 \ ,
    \label{eq:af}
\end{align}
characterizes the distribution of entanglement at a bipartition $AB$ in the system. 
High values of ${\cal F}_{AB}$ correspond to fewer significant components contributing to the entanglement distribution, whereas low ${\cal F}_{AB}$ indicates a more uniform entanglement structure.
Figure~\ref{fig:af_ee} shows ${\cal F}_{AB}$ and the entanglement entropy $S_{AB}$ at $n=L/2$ in late-time states of collision simulations as a function of initial momentum $k_i$.
The decrease in ${\cal F}_{AB}$ and increase in $S_{AB}$ past $k_i=k_\text{thr}$ indicates that there are more significant Schmidt components with increasing momentum.\footnote{The changes in ${\cal F}_{AB}$ and $S_{AB}$ are not sharp because the wavepackets have $\sigma_k\neq0$.}
This corresponds to the inelastic channel opening up.
Entanglement is maximized around $k_\text{thr}$ because there is significant probability to produce a particle $|2\rangle$ at rest or moving slowly, whose wavefunction is bisected by a bipartition at $L/2$. 
This relation of entanglement with the physical processes that drive the structure of the state can also be seen explicitly in Table~\ref{tab:schmidt_vals}, where the significant Schmidt components are shown for each $k_i$.
\begin{table}
\centering
\renewcommand{\arraystretch}{1.4}
\begin{tabularx}{\linewidth}{|c||Y|Y|Y|Y|Y|Y|} \hline
$k_i$ & $0.18\pi$ & $0.2\pi$ & $0.24\pi$ & $0.28\pi$ &
$0.32\pi$ & $0.36\pi$ \\\hline\hline
& & & & 0.6595, & 0.6574, & 0.6620,\\ 
$\lambda_i>10^{-1}$ & 0.8844 & 0.8325 & 0.6983 & 0.1296, & 0.1545, & 0.1570, \\
& & & & 0.1170 & 0.1376 & 0.1431 \\\hline
\end{tabularx}
\renewcommand{\arraystretch}{1}
\caption{{\it Significant components of the post-scattering state.} 
The dominant values of the $t=120$ entanglement spectrum $\{\lambda_i\}$ (Schmidt spectrum squared) at $L/2$ of $|f\rangle$ for various momenta $k_i$. 
}
\label{tab:schmidt_vals}
\end{table}

The entanglement structure at the center of the lattice reveals the composition of the late-time wavefunction, but does not distinguish the individual channels.
As explained in Sec.~\ref{sec:method}, a series of spatial Schmidt decompositions is used to distinguish the elastic from the inelastic channels. 
The cut locations are shown schematically in the top right panel of Fig.~\ref{fig:channel} and are chosen to maximize the separation of the elastic and inelastic components.

In general, $|1^{(12)}\rangle$ travels at a different velocity than $|1^{(11)}\rangle$ because some of its kinetic energy is converted to $|2^{(12)}\rangle$. 
However, an ultra high-energy collision, such as in the top left panel of Fig.~\ref{fig:channel}, leaves the velocity of $|1^{(12)}\rangle$ almost unchanged and its path largely overlaps with $|1^{(11)}\rangle$.
Compared to Fig.~\ref{fig:late_time_schematic} where there are three distinct regions on each half of the lattice, in Fig.~\ref{fig:channel} there are only two: ``fast'' and ``slow''.
The elastic channel is distinguished by having a particle in both fast regions, whereas $|12\rangle$ ($|21\rangle$) has a fast (slow) particle on the left and a slow (fast) particle on the right.
Since $\langle1^{(11)}|1^{(12)}\rangle\neq0$, the light particle in the inelastic channel can be decomposed as $|1^{(12)}\rangle = a|1^{(11)}\rangle + b|\delta\rangle$ where $\langle1^{(11)}|\delta\rangle=0$.
The Schmidt decomposition at $n_l$ diagonalizes the reduced density matrix of the left side giving three orthogonal left Schmidt vectors: $|1^{(11)}\rangle$, $|\delta\rangle$, and $|2^{(21)}\rangle$.
The late-time state in Eq.~\eqref{eq:production} becomes
\begin{equation}
\begin{aligned}
    |f \rangle \
&= \ | 1^{(11)} \rangle \otimes
\left(
\alpha | 1^{(11)} \rangle + a\beta | 2^{(12)} \rangle
\right) \\
&+ \ b \beta |\delta\rangle \otimes |2^{(12)}\rangle \\
&+ \ \beta |\text{vac}\rangle \otimes | 2^{(21)} \rangle | 1^{(21)} \rangle  \ .
\label{eq:cut_1}
\end{aligned}
\end{equation}
This separates the light left-moving components belonging to both channels from the heavy left-moving particle, which is identified as the last term in Eq.~\eqref{eq:cut_1}.

The right-moving $|1^{(11)}\rangle$ and $|2^{(12)}\rangle$ are orthogonal since they are different particle types at different positions. 
Since the $a$ and $b$ components of $|1^{(12)}\rangle$ share the same right-moving state $|2^{(12)}\rangle$, a second Schmidt decomposition at $n_r$ groups them back together.
This distinguishes the remaining $|11\rangle$ and $|12\rangle$ states.\footnote{Components containing the tails of the wavepackets due to finite particle separation and wavepacket spreading are suppressed, as are components corresponding to higher-order inelastic processes. See App.~\ref{app:channel_details} for details.}

The energy densities of the exclusive final states are shown in the lower panels of Fig.~\ref{fig:channel}.
In this figure, the difference between $v_1(k_i)$ and $v_1(k_f)$ is seen by comparing the peaks of the fast-moving components in the elastic and inelastic channels.
The broader profile of outgoing $|2\rangle$ particles results from larger variations in $v_2$ than $v_1$ around $k_f$.
Sharp features in the energy density are the result of bipartitions taken between wavepackets that have not fully separated.

The entanglement is found to be spread roughly equally between $|11\rangle$, $|12\rangle$, and $|21\rangle$. 
Each state has a single dominant Schmidt component (${\cal F}_{AB}\sim0$), consistent with unentangled asymptotic outgoing particles.\footnote{This split is similar in nature to ``configuration entropy'' and entropy within symmetry sectors in a gauge theory~\cite{Ghosh:2015iwa,Turkeshi:2020yxd}.}
Numerical values for ${\cal F}_{AB}$, $S_{AB}$, and the entanglement spectrum are given in App.~\ref{app:channel_details}.

\begin{figure}
    \centering
    \includegraphics[width=\linewidth]{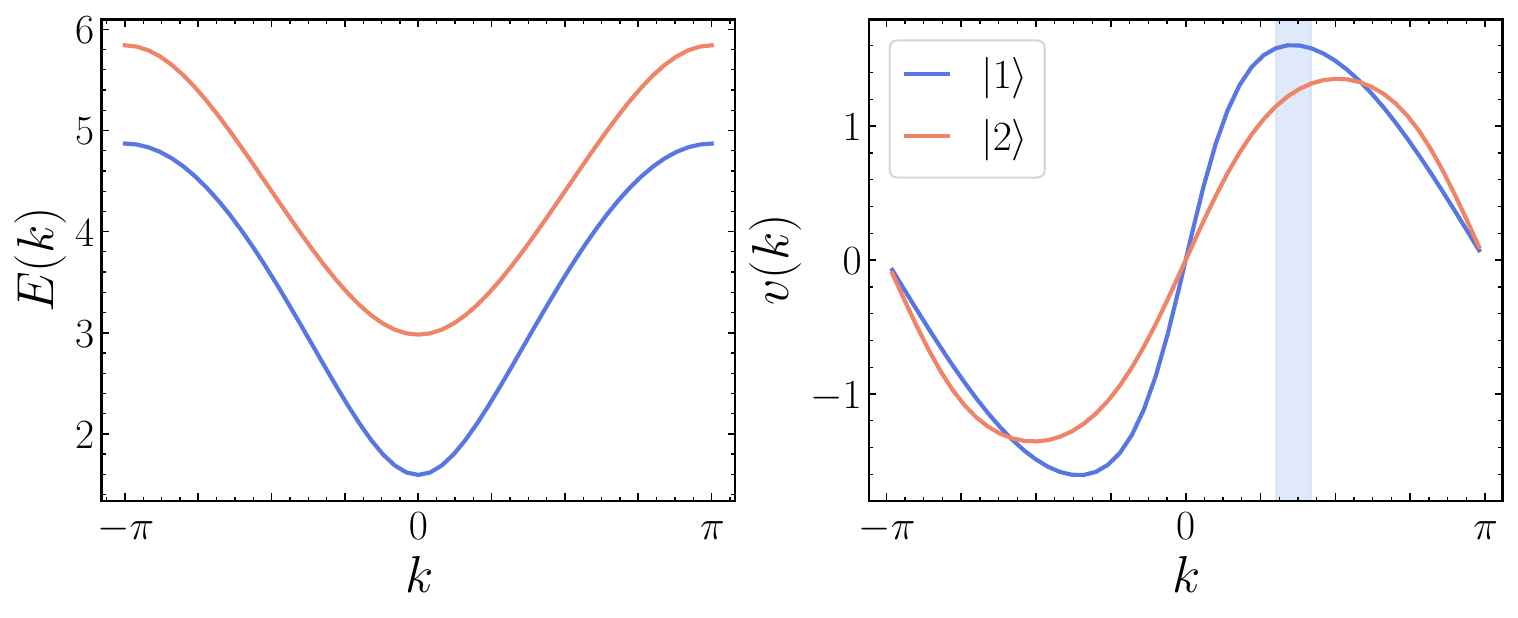}
    \caption{
    Dispersion relations $E(k)$ (left) and the group velocities $v(k)=dE(k)/dk$ (right) for particles $|1\rangle$ (blue) and $|2\rangle$ (orange) calculated in a $L=\infty$ system using the quasiparticle excitation ansatz.
    The light blue band on the right plot shows $k_i\pm\sigma_k$ used for the scattering simulations in this work.
    }
    \label{fig:dispersion}
\end{figure}

The particles within each state are then classified by mass using dispersion relations and group velocities calculated in the $L=\infty$ system, confirming the presence of a $|2\rangle$ particle (as opposed to, e.g., a slow $|1\rangle$ particle).
The group velocities $v(k)$ shown in the right panel of Fig.~\ref{fig:dispersion} extract the momenta $k(v)$ of particles traveling with speed $v$ measured by tracking peaks in $E_n$ over time.
The momentum is then used in the dispersion relation (left panel of Fig.~\ref{fig:dispersion}) to compute the energy $E(k(v))$ that particles $|1\rangle$ and $|2\rangle$ traveling with velocity $v$ would have.
Masses are determined by comparing $E(k(v))$ to the energy of the individual excitations $E_\text{wp}$ measured in each exclusive state and searching for (approximate) equality.\footnote{In addition to approximations discussed in Sec.~\ref{sec:method}, the use of infinite-system dispersion relations also introduces (exponentially-small) errors.}
These calculations are detailed in App.~\ref{app:particle_classification}.
Applying this self-consistent classification shows $E(k(v))$ and $E_\text{wp}$ are in agreement up to 8\% error, which can likely be improved with larger wavepackets and better spatial separation. 
This classification confirms the absence of $|2\rangle$ particles in the elastic channel and the presence of exactly one $|2\rangle$ particle in each inelastic state.

The branching ratios for $k_i=0.36\pi$ calculated from the decomposition are
\begin{align}
    P(11) \ = \ 0.5638\ , \quad P(12) \ = \ 0.3395 \ .
\end{align} 
Based on the identifications of the Schmidt components with the channels, 2-3 scattering ($11\to111$) and other higher-order inelastic processes have probabilities $O(10^{-2})$.
Up to approximations discussed in Sec.~\ref{sec:method}, these branching ratios are in agreement with previous results~\cite{Jha:2024jan} and could be confirmed by perturbative calculations near the integrable points of the theory. 
These calculations, together with the classification of particles in individual channels, confirm that a heavy $|2\rangle$ particle is produced in this collision event with probability 0.3395.

\section{Discussion}
\label{sec:conclusion}
\noindent
This technique provides a way to measure inelastic cross sections in simulations of scattering and is directly motivated by experimental detection protocols at collider facilities.
Akin to path reconstruction from layered detectors in collision experiments, spatial bipartitions separate wavefunction amplitudes that are then combined to form channel amplitudes, with particle velocities serving the role of time-of-flight measurements.
Notably, this method shows that a limited number of detectors, informed only by kinematics and symmetries, is enough to classify inelastic channels.
A promising direction for quantum simulations is to make this experimental analogy direct, implementing channel tagging by placing ``detectors'' throughout the lattice.
On each {\it shot} of the simulation, detector readings would indicate the presence or absence of a (certain type of) particle, allowing shot results to be directly categorized into channels.
Detectors in such simulations would parallel the role of bipartitions of MPS wavefunctions in this work.

While the utility of successive bipartitions is clear for cases where the kinematics constrains the outgoing momenta, higher-order events such as $11\to111$ will produce a distribution of states with different momenta, precluding the construction of, e.g., Dalitz plots~\cite{Dalitz:1953cp,Dalitz:1954cq} from spatial information alone.
This fundamental limitation of spatial cuts is encountered in 2-2 scattering if attempting to distinguish momentum modes within a wavepacket.
In such cases, the Schmidt spectrum still reflects the number of contributing channels, but isolating individual wavefunctions requires spatial separation of the outgoing particles.
A similar challenge exists in simulations of higher-dimensional systems, where event-shape observables such as Fox-Wolfram moments~\cite{Fox:1978vw} could provide a basis for channel classification beyond spatial bipartitions.
Addressing this limitation likely requires the incorporation of detectors both in classical and quantum simulations, with projections onto detector outcomes providing channel classification.

Spatial entanglement structure plays a crucial role in the channel isolation method and reveals the structure of the inclusive post-collision state. 
The direct correspondence of the entanglement spectrum with scattering channels points toward a potential broader relationship between entanglement and the S matrix~\cite{Vanderstraeten:2013xda,Beane:2018oxh,Beane:2021zvo,Beane:2020wjl}, which could be explored through the entanglement Hamiltonian~\cite{https://doi.org/10.1002/andp.202200064}.
While spatial bipartitions are natural for MPS, quantum simulations do not carry the restrictions associated with a spatial representation. 
As a result, this observable-driven measurement design generalizes to any quantum number that distinguishes outgoing channels, providing a framework for exclusive measurements in quantum simulations of scattering.

\section*{Acknowledgments}
\noindent
The author thanks Christian Bauer, Anthony Ciavarella, Ivan Burbano, Sarah Powell, Roland Farrell, Marc Illa, Henry Froland, Ivan Chernyshev, and Martin Savage for helpful discussions. 
This work was supported, in part, by U.S. Department of Energy, Office of Science, Office of Nuclear Physics, InQubator for Quantum Simulation (IQuS) under Award Number DOE (NP) Award DE-SC0020970 via the program on Quantum Horizons: QIS Research and Innovation for Nuclear Science.
This work was also supported, in part, by the Department of Physics and the College of Arts and Sciences at the University of Washington.
This work was enabled, in part, by the use of advanced computational, storage and networking infrastructure provided by the Hyak supercomputer system at the University of Washington.
The Quimb~\cite{gray2018quimb} and MPSKit~\cite{van_damme_2025_14719461} libraries were used for tensor network computations.

\clearpage
\onecolumngrid
\appendix

\section{Ising Field Theory}
\label{app:ising_details}
\noindent 
The Ising field theory in one spatial dimension emerges from the tilted-field Ising model (Eq.~\eqref{eq:h_ising}) in the scaling limit near its quantum critical point.
At criticality, the Ising conformal field theory (CFT) governs the physics of a massless free Majorana fermion.
The QFT near the critical point is parameterized by the dimensionless ratio
\begin{equation}
\eta_{\text{latt}} \ = \ \frac{g_x-1}{\vert g_z\vert^{8/15}} \ ,
\end{equation}
which arises from the relevant deformations in the CFT. 
A family of massive QFTs are obtained by tuning to criticality while holding $\eta_\text{latt}$ fixed.
The endpoints of the parameter space correspond to integrable theories: $\eta_\text{latt}=0$ yields the $E_8$ theory of Zamolodchikov~\cite{Zamolodchikov:1989fp,Zamolodchikov:1989hfa}, while  $\eta_\text{latt}=\infty$ describes a massive free fermion.
Integrability is broken away from these points, opening up inelastic scattering channels in which collisions can produce new particles.
Although the system is far from criticality for the couplings $g_x=1.25,\ g_z=0.15$ used in this work, features of a continuum theory are still present~\cite{Farrell:2025nkx}.
With these couplings, the theory supports two stable particles, $|1\rangle$ with $m_1=1.59$ and $|2\rangle$ with $m_2=2.97$.
The process $|11\rangle\to|12\rangle$ is kinematically allowed above $k_\text{thr}=0.2412\pi$.
More details on the Ising field theory can be found in Ref.~\cite{Jha:2024jan} and the thresholds for other inelastic processes for $g_x,\ g_z$ used in this work are given in Ref.~\cite{Farrell:2025nkx}.

\section{Channel Isolation Details}
\subsection{Entanglement Spectrum and Higher-Order Channels}
\label{app:channel_details}
\noindent
To accommodate finite simulation time and wavepacket spreading, as well as a small difference in the outgoing velocities of $|1^{(11)}\rangle$ and $|1^{(12)}\rangle$, the general method described in Sec.~\ref{sec:method} is modified.
At the bipartition of the system at lattice site $n_l=53$ (see Fig.~\ref{fig:channel}), the system has three significant Schmidt components, $\lambda_0,\ \lambda_2,$ and $ \lambda_2$.
Here $\lambda_{l,r}$ labels the index of the Schmidt component at successive cuts $n_l,\ n_r$.
The first row of Table~\ref{tab:schmidt_vals_nl_nr} shows the values of the Schmidt coefficients at this bipartition.
\begin{table}
\centering
\renewcommand{\arraystretch}{1.4}
\begin{tabularx}{\linewidth}{|c|c|c||Y|} \hline
$|\psi\rangle$ & $\langle\psi|\psi\rangle$ & Cut location & $\lambda_i>10^{-2}$
\\\hline\hline
$|f\rangle$ & 1 & $n_l$ & $\lambda_0=0.6129$, $\lambda_1=0.2074$, $\lambda_2=0.1463$\\\hline
$|f_{02}\rangle$ & 0.7592 & $n_r $ & $\lambda_{02,0}=0.5638$, $\lambda_{02,1}=0.1747$, $\lambda_{02,2}=0.0259$ \\\hline
$|f_{1}\rangle$ & 0.2074 & $n_r $ & $\lambda_{1,0}=0.1648$, $\lambda_{1,1}=0.0300$  \\\hline
$|\text{vac}\rangle$ & 1 & $n_l$ & 0.9786, 0.0105 \\\hline
$|\text{vac}_0\rangle$ & 0.9786 & $n_r$ & 0.9665, 0.0105\\\hline
\end{tabularx}
\renewcommand{\arraystretch}{1}
\caption{The entanglement spectrum (Schmidt coefficients squared) $\{\lambda_i\}$ in the post-scattering state with $k_i=0.36\pi$ after successive cuts at locations $n_l$, $n_r$.
The state is shown in the first column. 
Where present, $l$ in $|\psi_l\rangle$ corresponds to the components that are kept at the first bipartition. 
The second column shows the norm of the state in the first column. 
The fourth column gives the significant values of the entanglement spectrum obtained after cutting the state in the first column at the position given in third column.
}
\label{tab:schmidt_vals_nl_nr}
\end{table}
This cut separates out the heavy particles from the light particles moving left, but does not orthogonalize the light particles.
The wavefunction of each of these states is obtained by projecting onto the corresponding Schmidt component as described in Sec.~\ref{sec:method}.
Since $\langle1^{(11)}|1^{(21)}\rangle\neq0$, some mixing between $|11\rangle$ and $|12\rangle$ is present in the decomposition.
By inspecting the energy density of each wavefunction, it is determined that $|f_0\rangle$ and $|f_2\rangle$ mix the $|11\rangle$ and $|12\rangle$ states, while $|f_1\rangle$ contains $|21\rangle$. 

\begin{figure}
    \centering
\includegraphics[width=0.6\linewidth]{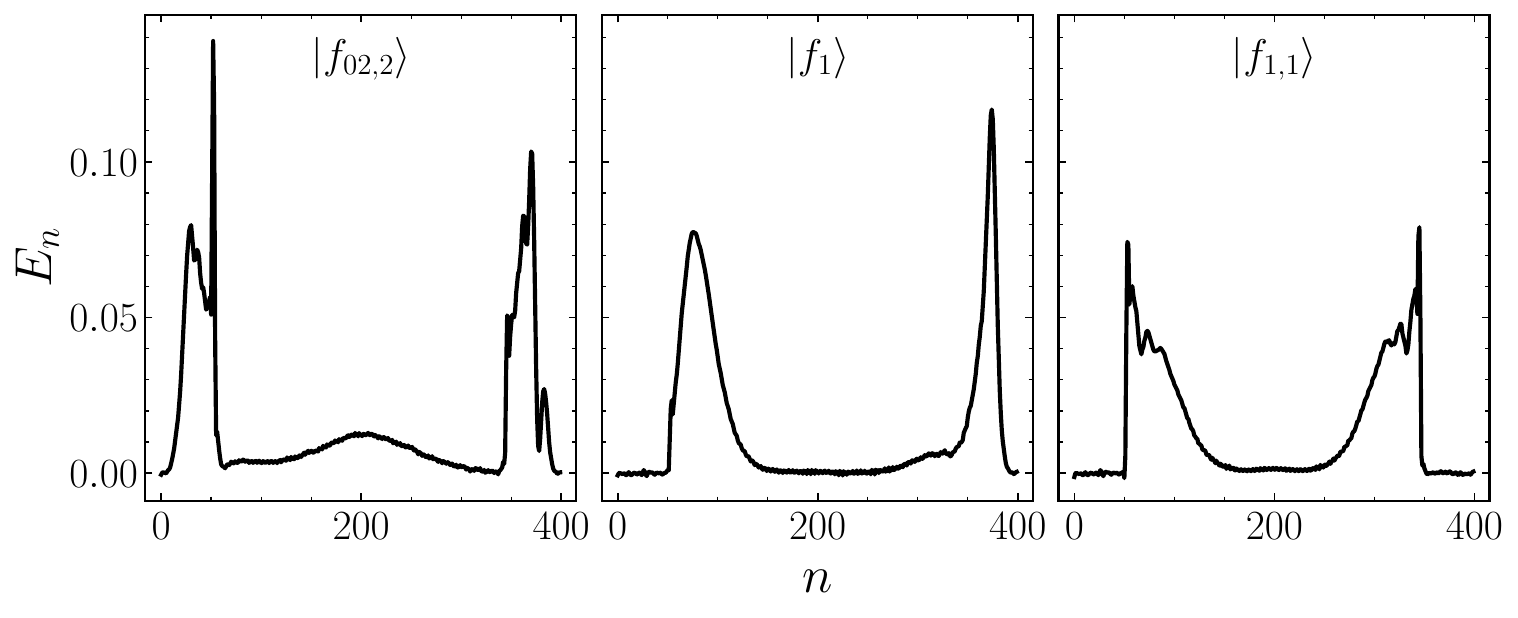}
\caption{
Energy densities $E_n$ for the Schmidt components used in the channel isolation process. 
The left panel shows $|f_{02,2}\rangle$, which is attributed to higher-order processes.
The middle panel shows $|f_1\rangle$, which is the combination of $|21\rangle$ and $|f_{1,1}\rangle$ shown on the right.
The state $|f_{1,1}\rangle$ is also attributed to higher-order processes.
}
\label{fig:other_channels}
\end{figure}

Projecting out the $|f_1\rangle$ retains the $|11\rangle$ and $|12\rangle$ states ($|f_{02}\rangle$), which are distinguished by a second bipartition at $n_r=L-n_l-1=346$.
Within $|f_{02}\rangle$, the second cut at $n_r$ also has three significant Schmidt coefficients: $\lambda_{02,0}$, which is identified with $|11\rangle$, $\lambda_{02,1}$, corresponding to $|12\rangle$, and $\lambda_{02,2}$.
The numerical values of these are given in the second row of Table~\ref{tab:schmidt_vals_nl_nr}.
The energy density in the last component, $|f_{02,2}\rangle$, is shown in the left panel of Fig.~\ref{fig:other_channels}, and is classified as a contribution from higher-order inelastic processes such as $11\to111$, as well as some potential mixing with the elastic channel.\footnote{Sharp features in the energy density around $n_l$ and $n_r$ are caused by forcing tensor product states at these locations.}
Parity symmetry requires the $|12\rangle$ and $|21\rangle$ components to be equal in amplitude, and rough equality is observed as expected ($\lambda_{02,1}\sim\lambda_{1,0}$).
Further decomposition of this wavefunction is only possible by cuts where excitations have significant support, which would separate different momentum modes belonging to the same channel, and are not done in this work.

The middle panel of Fig.~\ref{fig:other_channels} shows the energy density of the wavefunction $|f_1\rangle$ obtained by retaining the $\lambda_{1}$ component at the bipartition $n_l$, where a tail is seen on the left side of the $|1^{(21)}\rangle$ particle.
Performing a cut at $n_r$ on this wavefunction gives two significant Schmidt components $\lambda_{1,0}$ and $\lambda_{1,1}$, which are shown in the third row of Table~\ref{tab:schmidt_vals_nl_nr}.
The energy density of $|f_{1,0}\rangle$ is classified as $|21\rangle$ and is shown in the bottom right panel of Fig.~\ref{fig:channel}. 
The second component $|f_{1,1}\rangle$ is shown in the right panel of Fig.~\ref{fig:other_channels}, and is deemed to similarly arise from higher-order processes.
As a result, $\lambda_{02,2}+\lambda_{1,1}$ is reported as the contribution from higher-order channels.
Schmidt decompositions of the vacuum are dominated by a single vector across all cuts, shown in the bottom two rows of Table~\ref{tab:schmidt_vals_nl_nr}.
As a result of translation invariance, $|\text{vac}\rangle$ has the same entanglement structure throughout the lattice.

In summary, the states contributing to the elastic and inelastic channels are isolated by selecting the following Schmidt components:
\begin{equation}
    |f_{02,0}\rangle \ = \ |11\rangle \ , \ 
    |f_{02,1}\rangle \ = \ |12\rangle \ , \ 
    |f_{1,0}\rangle \ = \ |21\rangle
\end{equation}

Table~\ref{tab:ee_af_channel} shows the dominant Schmidt components, ${\cal F}_{AB}$ and $S_{AB}$ in the exclusive final states across a bipartition at the center of the lattice.
The remaining components are $O(10^{-3})$, suppressed by two orders of magnitude compared to the dominant states.
Together with the near-zero values of ${\cal F}_{AB}$, this indicates that the exclusive states $|11\rangle$, $|12\rangle$, and $|21\rangle$ are well described by product states, consistent with unentangled asymptotic outgoing particles.
The entanglement appears to be shared roughly evenly between the states, with variations likely caused by the differences in the cutting procedure described earlier.
Compared to $|12\rangle$ and $|21\rangle$, $|11\rangle$ is seen to have a larger value of ${\cal F}_{AB}$.
This is caused by the elastic channel having a single dominant component with more weight than the main dominant components in the inelastic channel. 
\begin{table}
\centering
\renewcommand{\arraystretch}{1.4}
\begin{tabularx}{\linewidth}{|c||Y|Y|Y|} \hline
State & $\lambda_i>10^{-2}$ & ${\cal F}_{AB}$ & $S_{AB}$
\\\hline\hline
$|f\rangle = |11\rangle + |12\rangle + |21\rangle$ & 0.6620,0.1570,0.1431 & $1.373\times 10^{-3}$ & 1.5380 \\ \hline\hline
$|f_{02,0}\rangle = |11\rangle$ & 0.5553 & $8.7843\times10^{-4}$ & 0.5419 \\ \hline
$|f_{02,1}\rangle = |12\rangle$ & 0.1621 & $7.4977\times10^{-5}$ & 0.5347 \\ \hline
$|f_{1,0}\rangle = |21\rangle$ & 0.1607 & $7.3546\times10^{-5}$ & 0.4646 \\ \hline
\end{tabularx}
\renewcommand{\arraystretch}{1}
\caption{Values of entanglement measures of the final states (first column) across a bipartition at $L/2$.
The second column gives significant values in the entanglement spectrum (Schmidt spectrum squared) at $t=120$.
The antiflatness ${\cal F}_{AB}$ and entanglement entropy $S_{AB}$ are given in the third and fourth columns respectively. 
The bottom three rows give the exclusive states determined from the inclusive state $|f\rangle$ (first row).
The exclusive states are not normalized so that their probabilities and entropies sum to the values in the first row.
}
\label{tab:ee_af_channel}
\end{table}
%

\subsection{Particle Classification}
\label{app:particle_classification}
\noindent
After the states contributing to $|f\rangle$ are isolated, the kinematics of the process can be used to classify the particles by type within each channel. 
Table~\ref{tab:disp_calc} shows the details of the classification calculation. 
The velocity $v$ of the excitations in each state is given in the second column.
It may be computed by isolating exclusive states at several times and comparing positions of peaks in $E_n$.
Another method to determine $v$ involves finding the collision time $t_0$ by maximizing $E_{L/2}(t)$ and finding the velocity starting of each excitation from $t_0$ to $t=120$.
Equivalently, $t_0$ can be determined by a calibration to the 11 channel where the outgoing speed is known, which gives $t_0=8.2$ (compared with $t_0=8$ obtained from maximizing $E_n$).
The velocity computed with the latter method is reported in Table~\ref{tab:disp_calc}.
The group velocity $v_{1,2}(k)$ is then inverted to determine $k_{1,2}(v)$, which is plugged into the dispersion relation to determine $E_{1,2}(k_{1,2}(v))$.
Where present, the several entries in a single column correspond to cases where multiple solutions are possible.
A $-$ entry in Table~\ref{tab:disp_calc} indicates there is no solution, eliminating the possibility to classify the given excitation as that type of particle.
A classification between $|1\rangle$ and $|2\rangle$ is made by comparing $E(k(v))$ to the energy of the individual excitations measured in each exclusive state $|\psi_\text{chan.}\rangle$, ${E_{\text{wp}} = \sum_{n\in\text{wp}} \langle \psi_\text{chan.}|\hat H_n |\psi_\text{chan.}\rangle -\langle \text{vac}|\hat H_n |\text{vac}\rangle}$ and selecting the closest match.
In all cases, the error $|E_\text{wp}-E(k(v))|/E_\text{wp}<8\%$ and the classification identifies the absence of $|2\rangle$ in the elastic channel and the presence of a single $|2\rangle$ in the inelastic states.
Making a misclassification requires a 28\% error in the calculation.
Mismatches in the energies determined in this process stem from the approximate nature of the isolation process, as well as approximate state preparation.
\begin{table}
\centering
\renewcommand{\arraystretch}{1.4}
\begin{tabularx}{\linewidth}{|c||Y|Y|Y|Y|Y|Y||c|} \hline
State & $v$ & $k_1(v)$ & $k_2(v)$ & $E_1(k_1(v))$ & $E_2(k_2(v))$ & $E_\text{wp}$ & Classification
\\\hline\hline
$|1^{(11)}\rangle$ & 1.6057 & 0.3551,\textbf{0.3600} & -- & 2.8489,\textbf{2.8735} & -- & 2.8512 & $|1\rangle$ \\\hline
$|1^{(12)}\rangle$ & -1.5610 & -0.4428,\textbf{-0.2817} & -- & 3.2877,\textbf{2.4816} & -- & 2.3616 & $|1\rangle$ \\\hline
$|2^{(12)}\rangle$ & 1.0958 & 0.1361,0.6771 & \textbf{0.2779},0.7338 & 1.8491,4.2918 & \textbf{3.5154},5.3364 & 3.3464 & $|2\rangle$ \\\hline
$|2^{(21)}\rangle$ & -1.0869 & -0.6803,-0.1346 & -0.7374,\textbf{-0.2742} & 4.3027,1.8437 & 5.3486,\textbf{3.5026} & 3.3527 & $|2\rangle$ \\\hline
$|1^{(21)}\rangle$ & 1.5610 & \textbf{0.2817},0.4428 & -- & \textbf{2.4816},3.2877 & -- & 2.3088 & $|1\rangle$ \\\hline
\end{tabularx}
\renewcommand{\arraystretch}{1}
\caption{{\it Particle classification based on measured energy and velocity of local observables.}
For a given individual excitation (first column) in the exclusive states, the second column gives the velocity of the peaks in $E_n$ over time.
The third and fourth columns give the results of extracting the possible momenta from the group velocities of particles $|1\rangle$ and $|2\rangle$ respectively. 
A $-$ entry indicates no solution is found, and multiple entries in a single column correspond to cases where multiple solutions are possible.
The fifth and sixth columns show the energy calculated from the dispersion relation using the momenta from the third and fourth columns.
The seventh column gives the energy $E_\text{wp}$ of the excitations calculated from the area under $E_n$ in the bottom panel of Fig.~\ref{fig:channel}.
The last column shows the classification of each excitation into particle type.
Bold entries correspond to the closest match between $E(k(v))$ (columns five and six) and $E_\text{wp}$ (column seven).
Since the left-moving $|1^{(11)}\rangle$ only differs from the right-moving $|1^{(11)}\rangle$ by the sign of its velocity, its calculations are not included in this table.
}
\label{tab:disp_calc}
\end{table}
%

\section{Momentum Measurement in Classical and Quantum Simulations}
\label{app:momentum}
\noindent
The simplest way to estimate the momentum of excitations traveling on the lattice is by recording the velocity of local observables sensitive to the excitations, such as $E_n$.
This is the method that is used in Sec.~\ref{sec:results} and has no extra overhead in quantum or classical simulations.
When $E(k)$ is now known, several other techniques are possible both in classical and classical quantum simulations.

In the context of MPS, momentum can be measured in real-time by overlapping with plane-wave states, such as those prepared by the quasiparticle excitation ansatz described in App.~\ref{app:sim_details}. 
In principle, any basis of states with well-defined momentum, such as those in Eq.~\eqref{eq:spatial_v}, can be used to perform this measurement.
This is efficient with MPS, and may be done by sampling a subset of momenta instead of measuring all $2^{n_Q}$ modes.
This approach has also been used to detect particles by overlapping with localized wavepackets~\cite{Jha:2024jan}.

On a quantum computer, the momentum could be measured in a local region of the lattice by applying a unitary Fourier transform $V$ on the lattice indices.
This is the equivalent method of computing overlaps with plane-wave states in MPS described above.
On a 4-qubit region of the lattice, $V$ carries out the following transformation.
\begin{equation}
\begin{aligned}
    &\text{Hamming weight 0} \, \left\{\,
    \begin{aligned}
        |0000\rangle \ &\to \ k=0: \ &&|0000\rangle & 
    \end{aligned}
    \right.&
    \\[10pt]
    &\text{Hamming weight 1} \left\{
    \begin{aligned}
        |0001\rangle \ &\to \ k=0: &&\frac{1}{2} \left( |0001\rangle + |0010\rangle + |0100\rangle + |1000\rangle \right) \\
        |0010\rangle \ &\to \ k=\frac{\pi}{2}: &&\frac{1}{2} \left( |0001\rangle + i|0010\rangle - |0100\rangle - i|1000\rangle \right) \\
        |0100\rangle \ &\to k=\pi: &&\frac{1}{2} \left( |0001\rangle - |0010\rangle + |0100\rangle - |1000\rangle \right) \\
        |1000\rangle \ &\to \ k=-\frac{\pi}{2}: &&\frac{1}{2} \left( |0001\rangle - i|0010\rangle - |0100\rangle + i|1000\rangle \right)
    \end{aligned}
    \right.&
    \\[10pt]
    &\text{Hamming weight 2} \left\{
    \begin{aligned}
        |0011\rangle \ &\to \ k=0: &&\frac{1}{2} \left( |0011\rangle + |0110\rangle + |1100\rangle + |1001\rangle \right) \\
        |0110\rangle \ &\to \ k=\frac{\pi}{2}: &&\frac{1}{2} \left( |0011\rangle + i|0110\rangle - |1100\rangle - i|1001\rangle \right) \\
        |1100\rangle \ &\to \ k=\pi: &&\frac{1}{2} \left( |0011\rangle - |0110\rangle + |1100\rangle - |1001\rangle \right) \\
        |1001\rangle \ &\to \ k=-\frac{\pi}{2}: &&\frac{1}{2} \left( |0011\rangle - i|0110\rangle - |1100\rangle + i|1001\rangle \right) \\
        |0101\rangle \ &\to \ k=0: &&\frac{1}{\sqrt 2} \left( |0101\rangle + |1010\rangle \right) \\
        |1010\rangle \ &\to \ k=\pi: &&\frac{1}{\sqrt 2} \left( |0101\rangle - |1010\rangle \right)
    \end{aligned}
    \right.&
    \\[10pt]
    &\text{Hamming weight 3} \left\{
    \begin{aligned}
        |0111\rangle \ &\to \ k=0: &&\frac{1}{2} \left( |0111\rangle + |1110\rangle + |1101\rangle + |1011\rangle \right) \\
        |1110\rangle \ &\to \ k=\frac{\pi}{2}: &&\frac{1}{2} \left( |0111\rangle + i|1110\rangle - |1101\rangle - i|1011\rangle \right) \\
        |1101\rangle \ &\to \ k=\pi: &&\frac{1}{2} \left( |0111\rangle - |1110\rangle + |1101\rangle - |1011\rangle \right) \\
        |1011\rangle \ &\to \ k=-\frac{\pi}{2}: &&\frac{1}{2} \left( |0111\rangle - i|1110\rangle - |1101\rangle + i|1011\rangle \right)
    \end{aligned}
    \right.&
    \\[10pt]
    &\text{Hamming weight 4} \,\left\{\,
    \begin{aligned}
        |1111\rangle \ &\to \ &k=0: \ &&|1111\rangle
    \end{aligned}
    \right.&
\end{aligned}
\label{eq:spatial_v}
\end{equation}
While finding a circuit for this transformation is challenging, its structure would mirror the classical Cooley-Tukey Fast Fourier Transform (FFT)~\cite{Cooley:1965zz} on $n_Q=2^n$ qubits.
Similar circuits have been developed for the fermionic FFT~\cite{Ferris:2014jmb,Barak:07,Verstraete:2008qpa}, and have been used in the context of neutral atom platforms~\cite{Maskara:2025oab} to implement fermion permutations.
Following the transformation to (spatial) momentum space (in contrast to Hilbert space momentum), measurements on the qubits within the ``detector'' region would correspond to the momentum modes present in that region.
This would be an approximate measurement of momentum with the resolution set by the size of the detector $L_d$, yielding a momentum uncertainty $\Delta k\sim 2\pi/L_d$ at the cost of a circuit depth overhead.

\begin{figure}
    \centering
\includegraphics[width=0.9\linewidth]{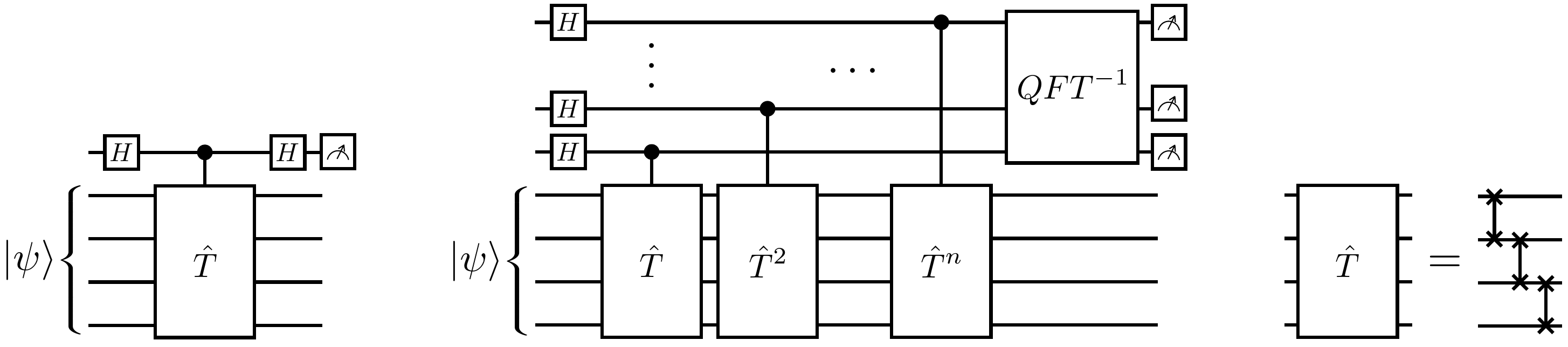}
\caption{
Circuits that could be used to measure momentum using the translation operator $\hat T$.
Left: the circuit that measures $\langle \psi|\hat T|\psi\rangle$. 
Center: the quantum phase estimation circuit that can be used for momentum measurement.
Right: the implementation of the translation operator in a spin system.
}
\label{fig:measure_t}
\end{figure}

Another approach involves measuring the expectation value of the translation operator $\hat T$. 
On a 4-qubit state $\hat T$ acts as $\hat T|0001\rangle \ = \ |0010\rangle$.
On a state with well-defined momentum $|\psi_k\rangle$, $\hat T$ shifts the state by one lattice site and returns a phase 
\begin{align}
    \hat T|k\rangle \ = \ e^{ik}|k\rangle \ .
    \label{eq:T_action}
\end{align}
The expectation value $\langle \hat T\rangle$ can be measured using the Hadamard test~\cite{10.1098/rspa.1998.0164}, requiring a single application of a controlled-$\hat T$ unitary.
The circuit implementing this is shown in the left panel of Fig.~\ref{fig:measure_t}.
The probability of measuring 0 on the ancilla qubit is related to $\langle \hat T\rangle$ by $P(0) = \frac{1+Re\langle \hat T\rangle}{2}$.
The imaginary component of the phase is similarly extracted by measuring in the $Y$ basis.
For spin systems, this circuit requires a network of controlled-swap gates, and is long-range in nature.
This approach may also be extended to full phase estimation, requiring additional ancilla qubits and applications of the controlled-$\hat T$ unitary. 
This is shown in the central panel of Fig.~\ref{fig:measure_t}, and would give information regarding the amplitude present in each momentum sector, up to the resolution set by the number of ancilla qubits.\footnote{Recently, there have been optimizations to the phase estimation algorithm developed which allow full phase estimation with a single ancilla qubit at the expense of sample overhead, and removing controlled-unitary dependence altogether~\cite{Clinton:2024sij}.}
This approach carries an overhead in shots and circuit depth.
Since proper momentum states only have the phase relation of Eq.~\eqref{eq:T_action} when translated along the whole lattice, this only yields approximate measurements of momentum when applied locally, and is thus of limited use.
However, it could be used to test for local translation invariance.

\section{Computational Methods}
\label{app:sim_details}
\subsection{State Preparation}
\label{app:state_prep}
\begin{figure}
    \centering
\includegraphics[width=0.5\linewidth]{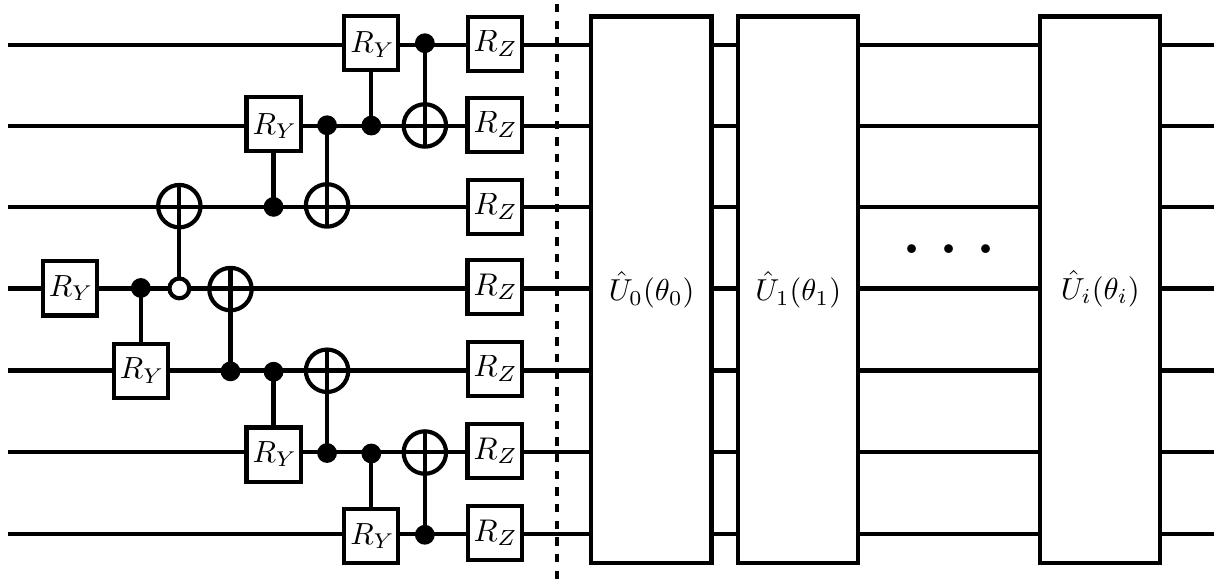}
\caption{
The circuit for preparing $|\psi_\text{wp}\rangle$ in Eq.~\eqref{eq:wf_wp}.
The left side of the circuit prepares the initial state $|W(k_i, \sigma_k)\rangle$. 
The $R_Y$ angles are obtained by solving Eq.~\eqref{eq:WP0angles}, and the $R_Z$ angles are determined from the phases $e^{-ikn_0}$ in Eq.~\eqref{eq:wf_wp}.
The right side of the circuit implements layers of translationally invariant energy minimization unitaries $\hat U_i(\theta_i)$ determined using ADAPT-VQE. 
}
\label{fig:wp_prep_circ}
\end{figure}

\noindent
The initial states of the scattering simulations in this work contain two $|1\rangle$ particles with momenta $\pm k_i$.
These particles are represented by localized wavepackets built out of plane-wave excitations $|\psi_k\rangle$ centered around $\pm k_i$ and lattice position $n_0$,
\begin{equation}
\vert \psi_{\text{wp}} \rangle  \ = \ {\cal N}\sum_k e^{-i k n_0}\, e^{-(k_i - k)^2/(2\sigma_k^2)} \vert \psi_k \rangle \ ,
\label{eq:wf_wp}
\end{equation}
where ${\cal N}$ is a normalization factor and $\sigma_k$ is the width in momentum space.
The state preparation algorithm proceeds in two steps. 
First, a state $|W(k_i,\sigma_k)\rangle$ is initialized that establishes the momentum content, position, and quantum numbers of the target $|\psi_\text{wp}\rangle$, but has contributions from higher-energy states and multiple particles.
The circuit preparing $|W(k_i,\sigma_k)\rangle$ is shown in the left side of Fig.~\ref{fig:wp_prep_circ}. 
The $R_Y$ rotation angles are obtained by recursively solving the equations
\begin{align}
\left [\sin\left (\frac{\theta_{\eta}}{2}\right )\right ]^2 \ &= \ \sum_{i=\eta}^{d-1} \, c_i^2  \ , \nonumber \\[4pt]
\cos{\left (\frac{\theta_{\eta+j+1}}{2}\right )}\prod_{i=0}^{j} \sin{\left (\frac{\theta_{\eta+i} }{2}\right )} \ &= \ c_{\eta+j} \ \ , \ \ j\in[0,1,\ldots,\eta-1] \ , \nonumber \\[4pt] 
\cos{\left (\frac{\theta_{\eta}}{2}\right )}\cos{\left (\frac{\theta_{\eta-j-1}}{2}\right )}\prod_{i=2}^{j} \sin{\left (\frac{\theta_{\eta-i} }{2}\right )} \ &= \ c_{\eta-j} \ \ , \ \ j\in[1,2,\ldots,\eta-1] \ ,
\label{eq:WP0angles}
\end{align}
and the $R_Z$ angles are determined from the phases $e^{-ikn_0}$ in Eq.~\eqref{eq:wf_wp}.
For $\sigma_k=0$, these states correspond to plane waves of the form 
\begin{align}
    |k_i\rangle \ = \ \frac{1}{\sqrt{N}}\sum_n e^{ik_in} |2^n\rangle \ ,
\end{align}
which are exact single-particle plane waves in the limit $g_x=0$.
Here $|2^n\rangle$ represents the little-endian bitstring of a state with a single spin flipped at position $n$ (e.g., for a system of 4 spins, $|2^2\rangle = |0100\rangle$).
At $k_i=0,\ \sigma_k=0$, this state is a W state~\cite{Dur:2000zz}, which is widely studied in the context of quantum information for its entanglement properties.

The state $|W(k_i,\sigma_k)\rangle$ establishes the long-range entanglement present in the target $|\psi_\text{wp}\rangle$. 
Next, single-particle wavepackets for general $g_x,\ g_z$ are prepared by minimizing the energy within each momentum sector.
By projecting each momentum sector to its ground state, a single-particle wavepacket of $|1\rangle$ is formed.\footnote{This method assumes the momentum of the wavepacket is sufficiently far from 0.}
The variational method ADAPT-VQE~\cite{Grimsley:2018wnd} is used to optimize parameterized circuits $\hat U(\vec{\theta})=\hat U_{n-1}(\theta_{n-1})\hat U_{n-2}(\theta_{n-2})\dots \hat U_0(\theta_0)$ where $\hat U_i=e^{i\theta_i \hat O_i}$ that minimize the energy. 
The operators $\hat O_i$ are selected from a pool of operators $\{\hat O_i\}$ generated from the algebra of the Hamiltonian,
\begin{equation}
\begin{aligned}
\{ \hat O_i\} \ &= \ i\sum_n[\hat{H}, \hat{X}_n ] \ \cup \ i\sum_n [ \hat{H}, [ \hat{H}, [\hat{H}, \hat{X}_n ] ] ]\\
 &= \ \sum_{n=0}^{L-1} \bigg \{ \hat{Y}_n 
 ,   \hat{Z}_n\hat{Y}_{n+1}\hat{Z}_{n+2} , \left (\hat{Y}_n\hat{Z}_{n+1}+ \hat{Z}_n\hat{Y}_{n+1}\right )  ,   \left (\hat{Y}_n\hat{X}_{n+1}+ \hat{X}_n\hat{Y}_{n+1}\right )  ,  \left (\hat{Z}_n\hat{X}_{n+1}\hat{Y}_{n+2}+ \hat{Y}_n\hat{X}_{n+1}\hat{Z}_{n+2}\right )  \bigg \} \ .
\label{eq:opPool}
\end{aligned}
\end{equation}
By incorporating the symmetries of the Hamiltonian, this translationally invariant pool ensures $U(\vec\theta)$ does not change the momentum content, spatial position, or quantum numbers of $|W(k_i,\sigma_k)\rangle$.
The specific operators and angles used for $k_i=0.36\pi$ are given in Table~\ref{tab:adapt_vqe_params}.
\begin{table}
\renewcommand{\arraystretch}{1.4}
\begin{tabularx}{\linewidth}{|c || Y | Y | Y | Y | Y | Y | Y | Y |}
  \hline
 $\hat O_i$ & 
 $\hat{Y}$&
 $\hat{Y}\hat{Z}$  & 
 $\hat{Y}$ &  
 $\hat{Z}\hat{X}\hat{Y}$&   
 $\hat{Y}\hat{Z}$&  
 $\hat{Y}\hat{Z}$
 &
 $\hat{Y}$
&
 $\hat{Z}\hat{Y}\hat{Z}$
\\
 \hline\hline
$\theta_i$ &  0.1212 & 0.0185  & -0.5452 &  0.0397 & 0.0599& 0.0556 &-0.2637& 0.0566\\\hline
\end{tabularx}
\renewcommand{\arraystretch}{1}
\caption{
The operators and parameters used to initialize wavepackets with $k_i=0.36\pi$ and $\sigma_k=0.059\pi$. 
These are determined from a $L=256$ system using a MPS circuit simulator.
For simplicity, $\hat O_i$ is labeled by a single term, i.e., $\hat Y\hat Z$ corresponds to the operator $\sum_n^{L-1} \left(\hat Y_n \hat Z_{n+1} + \hat Z_n \hat Y_{n+1}\right)$.}
 \label{tab:adapt_vqe_params}
\end{table}
Circuit training is done on a $L=256$ system with an MPS circuit simulator.
This adjusts the short-range entanglement to match that of $|\psi_\text{wp}\rangle$.
See Ref.~\cite{Farrell:2024fit} for a full discussion of the state preparation method, as well as circuit optimizations for quantum hardware and choices of $\hat O_i$, $\theta_i$ for other momenta.

This state preparation algorithm initializes the localized wavepackets in Eq.~\eqref{eq:wf_wp} and the vacuum of the periodic boundary conditions (PBC) system elsewhere.
Although MPS are much more efficient for open boundary conditions (OBCs)~\cite{PhysRevLett.93.227205}, PBCs are used in this work to avoid propagating excitations from imperfect state preparation near the boundary in OBCs.

\begin{figure}
    \centering
\includegraphics[width=0.4\linewidth]{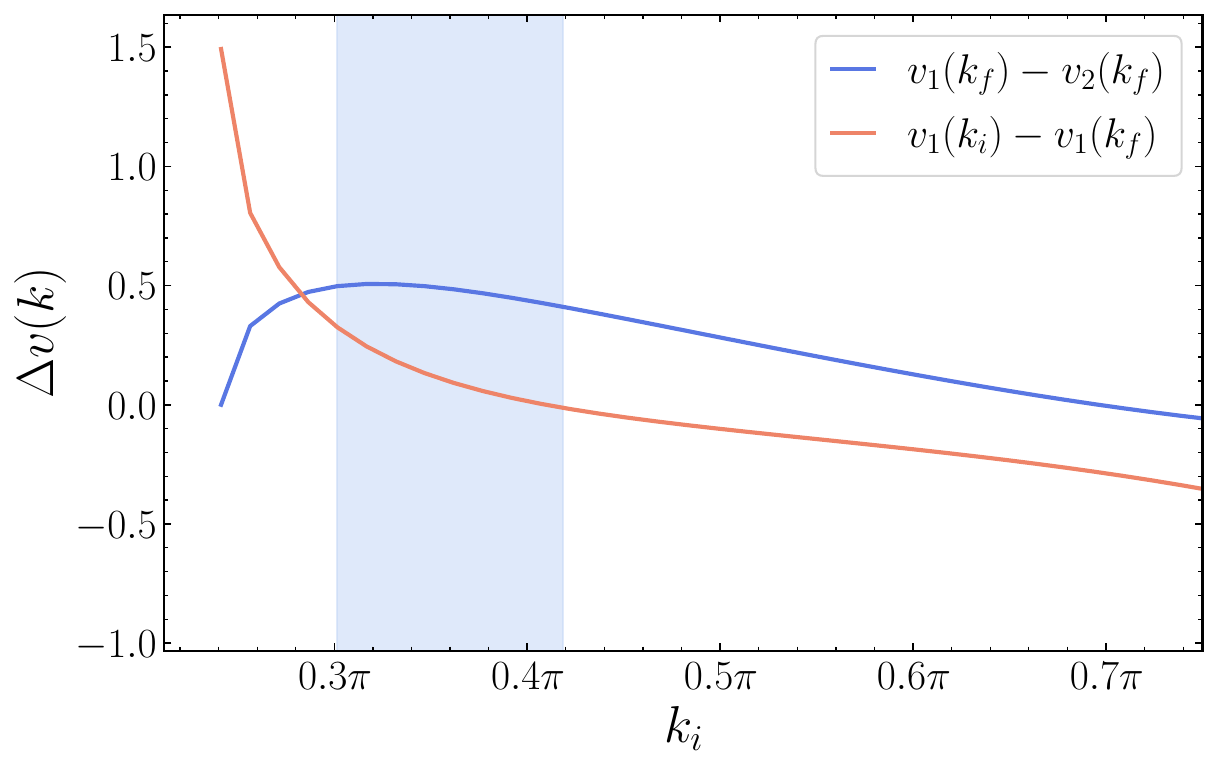}
\caption{
Velocity differences $\Delta v(k)$ of the outgoing particles as a function of incoming momentum $k_i$ of the $|1\rangle$ particles. 
The difference in velocity of $|1^{(12)}\rangle$ and $|2^{(21)}\rangle$ is shown in blue, while the difference between the velocities of the two light particles $|1^{(11)}\rangle$ and $|1^{(12)}\rangle$ is shown in orange.
The light blue shading shows $k_i\pm\sigma_k$ used in scattering simulations in Sec.~\ref{sec:results}.
}
\label{fig:outgoing_group_velocities}
\end{figure}

The initial momentum of the wavepackets $k_i=\pm0.36\pi$ is chosen for the simulations producing the final state shown in Fig.~\ref{fig:channel}.
Figure~\ref{fig:outgoing_group_velocities} shows the differences between the velocities of the outgoing particles in the elastic and inelastic channels, which set the spacing between the particles in Fig.~\ref{fig:late_time_schematic}.
While the separation between $|1^{(11)}\rangle$ and $|1^{(12)}\rangle$ is small at $k_i=0.36\pi$ and $t=120$, the difference in the outgoing velocities of the light and heavy particles is close to maximal.
A smaller $k_i$ would increase the difference $v_1(k_i)-v_1(k_f)$, but would decrease the velocities of all outgoing particles (see left panel of Fig.~\ref{fig:dispersion}).
The combined effects of wavepacket spreading and slower propagation result in the best spatial separation between $|1\rangle$ and $|2\rangle$ being near $k_i=0.36\pi$.

The wavepacket width is set to $\sigma_k=0.059\pi$ and 0.9954\% of the $|\psi_\text{wp}\rangle$ norm is retained by truncating the wavepacket in position space to $d=21$ sites.

\subsection{MPS Details}
\noindent 
The MPS time evolution simulations are done with a bond dimension of {\tt max\_bond} $=600$ and cutoff of $10^{-9}$ to $t=40$, well past the collision time ($t\sim8$).
After this time, the MPS is truncated to {\tt max\_bond} $=350$ with the same cutoff to increase the speed of the simulation, which introduces truncation errors of $O(10^{-3})$.
The norm of the state is used to estimate the truncation effects, and is shown for various times in the simulation in Table~\ref{tab:trunc_errors}.
\begin{table}
\centering
\renewcommand{\arraystretch}{1.4}
\begin{tabularx}{\linewidth}{|c||Y|Y|Y|Y|Y|Y|} \hline
& \multicolumn{6}{c|}{$\langle f|f\rangle$} \\ \hline
$k_i$ & $0.18\pi$ & $0.2\pi$ & $0.24\pi$ & $0.28\pi$ &
$0.32\pi$ & $0.36\pi$ \\\hline\hline
$t=40$, {\tt max\_bond}$=600$ & 0.9938 & 0.9945 & 0.9926 & 0.9914 & 0.9911 & 0.9928 \\\hline
$t=40$, {\tt max\_bond}$=350$ & 0.9908 & 0.9920 & 0.9886 & 0.9870 & 0.9868 & 0.9897 \\\hline
$t=120$, {\tt max\_bond}$=350$ & 0.9331 & 0.9546 & 0.9379 & 0.9313 & 0.9355 & 0.9488 \\\hline
\end{tabularx}
\renewcommand{\arraystretch}{1}
\caption{Truncation errors given by the norms of $|f\rangle$ at various times and initial momenta $k_i$ in the MPS simulation. The first row shows the norm well after the scattering events at $t=40$ at {\tt max\_bond}$=600$. 
At this point, {\tt max\_bond} is truncated to 350 (second row).
The third row shows the norms at the final time $t=230$.
All rows use a cutoff of $10^{-9}$.
}
\label{tab:trunc_errors}
\end{table}
The final states at $t=120$ have truncation errors of $<7\%$ and are normalized to unity for purposes of channel isolation in Sec.~\ref{sec:results}.
While this error is non-negligible, it is assumed that the most physically-important components are retained by the SVD truncation, since the $t=40$ post scattering state with {\tt max\_bond}$=600$ has norm near unity and no additional interactions take place after this time.
A trotter step size of $\delta t=1/32$ is used to simulate the scattering process to $t=120$. 
This could be improved by using an optimized higher-order Trotter formula~\cite{Barthel:2019kch}.\footnote{It is interesting to consider the possibility of simulating scattering events in momentum space, where the states are much simpler at the expense of having highly-nonlocal interactions~\cite{VanDamme:2022lax,Corbett:2025flm}.}

For one-dimensional gapped theories with isolated single-particle excitations, dispersion relations can be efficiently computed in MPS using the quasiparticle excitation ansatz~\cite{Haegeman:2013xcv}. 
Where this is not possible, they could be computed with exact diagonalization for small system sizes and extrapolated to suitable $L$~\cite{Farrell:2023fgd,Zemlevskiy:2024vxt}.
Dispersion relations could be evaluated on a quantum computers with knowledge of the single-particle excitation structure. 
This can be done by preparing single-particle plane waves (e.g., Ref.~\cite{Farrell:2025nkx}) and measuring their energy, or by preparing wavepackets and recording their speed.
The dispersion relations in Sec.\ref{sec:results} were constructed using with the quasiparticle excitation ansatz in the thermodynamic limit of the Hamiltonian in Eq.~\eqref{eq:h_ising}.
This algorithm first finds the (translationally invariant) vacuum of the $L=\infty$ system using Variational Uniform Matrix Product States (VUMPS)~\cite{Zauner-Stauber:2017eqw} specified by the tensor $A$,
\begin{equation}
|\text{vac}\rangle \ = \ \dots
\begin{tikzpicture}[baseline=-0.5ex, scale=0.8,
    tensor/.style={draw, rounded corners=3pt, minimum size=0.7cm, fill=white, inner sep=2pt}
]
    \node[tensor] (A0) at (0, 0) {$A$};
    \node[tensor] (A1) at (1.4, 0) {$A$};
    \node[tensor] (A)  at (2.8, 0) {$A$};
    \node[tensor] (A2) at (4.2, 0) {$A$};
    \node[tensor] (A3) at (5.6, 0) {$A$};

    \draw (A0.west) -- ++(-0.35,0);
    \draw (A0.east) -- (A1.west);
    \draw (A1.east) -- (A.west);
    \draw (A.east) -- (A2.west);
    \draw (A2.east) -- (A3.west);
    \draw (A3.east) -- ++(0.35,0);

    \foreach \t in {A0,A1,A,A2,A3} {
        \draw (\t.south) -- ++(0,-0.25);
    }

    \node[below] at ($(A0.south)+(0,-0.25)$) {\small $\cdots$};
    \node[below] at ($(A1.south)+(0,-0.25)$) {\small $s_{n-1}$};
    \node[below] at ($(A.south)+(0,-0.25)$)  {\small $s_n$};
    \node[below] at ($(A2.south)+(0,-0.25)$) {\small $s_{n+1}$};
    \node[below] at ($(A3.south)+(0,-0.25)$) {\small $\cdots$};
\end{tikzpicture}
\dots \ ,
\end{equation}
where the tensor legs below the tensors $A$ correspond to the physical indices of the state.
Low-lying excitations of momentum $k$ are constructed as momentum superpositions of a local operator $B_{k;j}$ acting on the vacuum,
\begin{equation}
|\psi_{k;j}\rangle \ = \ \sum_n e^{ikn} \dots
\begin{tikzpicture}[baseline=-0.5ex, scale=0.8,
    tensor/.style={draw, rounded corners=3pt, minimum size=0.7cm, fill=white, inner sep=2pt}
]
    \node[tensor] (A0) at (0, 0) {$A$};
    \node[tensor] (A1) at (1.4, 0) {$A$};
    \node[tensor] (B)  at (2.8, 0) {$B_{k;j}$};
    \node[tensor] (A2) at (4.2, 0) {$A$};
    \node[tensor] (A3) at (5.6, 0) {$A$};

    \draw (A0.west) -- ++(-0.35,0);
    \draw (A0.east) -- (A1.west);
    \draw (A1.east) -- (B.west);
    \draw (B.east) -- (A2.west);
    \draw (A2.east) -- (A3.west);
    \draw (A3.east) -- ++(0.35,0);

    \foreach \t in {A0,A1,B,A2,A3} {
        \draw (\t.south) -- ++(0,-0.25);
    }

    \node[below] at ($(A0.south)+(0,-0.25)$) {\small $\cdots$};
    \node[below] at ($(A1.south)+(0,-0.25)$) {\small $s_{n-1}$};
    \node[below] at ($(B.south)+(0,-0.25)$)  {\small $s_n$};
    \node[below] at ($(A2.south)+(0,-0.25)$) {\small $s_{n+1}$};
    \node[below] at ($(A3.south)+(0,-0.25)$) {\small $\cdots$};
\end{tikzpicture}
\dots \ .
\label{eq:psi_k_mps}
\end{equation}
Here $j$ specifies the excitation number, corresponding to particle $|1\rangle$, $|2\rangle$ or higher excitations. 
For the simulations in this work, the states $|\psi_{k;1}\rangle$ correspond to the single-particle plane waves discussed in App.~\ref{app:state_prep}.
This ansatz provides an approximation to the true momentum states of the theory with an error that is exponentially small in the support of the local operator $B_{k;j}$. 
The vacuum of the $L=\infty$ system is well-represented by a bond dimension of {\tt max\_bond}$=12$, and the dispersion relation computations require {\tt max\_bond}$=24$.

\bibliography{bibi}

\end{document}